\documentclass[twocolumn,aps,prd,amsmath,amssymb,nofootinbib]{revtex4-1}

\pdfoutput=1

\usepackage{graphicx,amsmath,amssymb,feynmf}

\newcommand{\nn}{\nonumber} 
\newcommand{\bea}{\begin{eqnarray}}
\newcommand{\eea}{\end{eqnarray}}
\newcommand{\beq}{\begin{equation}}
\newcommand{\eeq}{\end{equation}}

\newcommand{\cL}{\mathcal{L}}
\newcommand{\etan}{\eta_n}

\newcommand{\cM}{\mathcal{M}}
\newcommand{\g}{g}
\newcommand{\cg}{c_g}
\newcommand{\cJ}{\mathcal{J}}
\newcommand{\en}{e_n}
\newcommand{\qn}{Q_n}

\newcommand{\eps}{\epsilon}

\newcommand{\lm}{\lambda}
\newcommand{\LP}{L}
\newcommand{\ta}{\mathcal{T}}
\newcommand{\al}{a}
\newcommand{\ba}{b}
\newcommand{\Proj}{r}
\newcommand{\md}{\sigma}

\begin{document}

\title{Special Relativity from Soft Gravitons}

\author{
Mark P.  Hertzberg, McCullen Sandora}
\affiliation{Institute of Cosmology, Department of Physics and Astronomy,\\ 
Tufts University, Medford, MA 02155, USA}

\begin{abstract}
We study all translationally and rotationally invariant local theories involving massless spin 2 and spin 1 particles that mediate long range forces, allowing for general energy relations and violation of boost invariance. Although gauge invariance is not a priori required to describe non Lorentz invariant theories, we first establish that locality requires `soft gauge invariance'. Then by taking the soft graviton limit in scattering amplitudes, we prove that in addition to the usual requirement of universal graviton couplings, the  special relativistic energy-momentum relation is also required and must be exact. We contrast this to the case of theories with only spin $\leq1$ particles, where, although we can still derive charge conservation from locality, special relativity can be easily violated. We provide indications that the entire structure of relativity can be built up from spin 2 in this fashion. 
\end{abstract}

\maketitle

{\em Introduction}:---
Special relativity is in spectacular agreement with all current observations. It provides a beautiful unification of space and time. Combined with quantum mechanics, it explains much of what we observe in the world. When applied to massless spin 1 particles, it describes the structure of the electroweak and strong forces, and when applied to massless spin 2 particles, it describes the structure of gravitation.

However, one may enquire what is the {\em origin} of special relativity, or more specifically, the Lorentz symmetry; is it a structure that can be deformed easily at low energies; is it an accidental symmetry that must be exact at low energies but might be violated at high energies? In the Standard Model of particle physics, Colladay and Kostelecky \cite{Colladay:1998fq} and Coleman and Glashow \cite{Coleman:1998ti} found that one can easily deform the Lorentz symmetry. In fact 46 new CPT even couplings at the dimension 4 level are allowed, without affecting the unitarity of the theory, the degree of freedom counting, or leading to any known pathology. Of course such deformations are highly constrained by experiment \cite{Mattingly:2005re}, but it is interesting that it can be done so easily in theories with only spin $s\leq1$ particles.

In this Letter we would like to point out the tremendous theoretical difficulty of violating the Lorentz symmetry when massless spin 2 is included. In particular, we will allow our matter species and massless spin 2 particle to carry {\em any} dispersion relation, arbitrary violations of boost invariance, and we will prove using Weinberg's method \cite{Weinberg:1964ew} of demanding consistent soft graviton scattering that the special relativistic dispersion relation is required. We provide indications that the full structure of relativity can be built up too. Our only assumptions will be that we have translational and rotational invariance in some frame and that interactions avoid instantaneous action at a distance.

{\em Spin}:--
Compatibility with quantum mechanics and rotation invariance demands that particles transform under a unitary representation of the rotation group $SO(3)$. These representations are organized by spin in the usual fashion $s=0,\,1/2,\,1,\,3/2,\,2,\ldots$. Furthermore, there are two distinct classes of representations for particles with spin $s\geq 1$: (i) One class of representations, usually called ``massive", fill out the full set of spins along, say, the $z$-axis as $s_z=-s,-s+1,\ldots,s-1,s$, i.e., all $2s+1$ components of angular momentum. (ii) Another class of representations, usually called ``massless", only fill out the spins parallel to the direction of motion, known as helicity with $h=\pm s$. We note that the presence of these classes do not rely upon the presence of the Lorentz symmetry, but only the rotation symmetry. Without boost invariance, the ``massive" representations do transform in a complicated fashion, while the ``massless" representations transform in a relatively simple fashion. Note that neither representation is a priori gapped or gapless.

The quantum state of a massless spin 2 particle is specified by its momenta ${\bf q}$ and helicity $h$ as $|{\bf q},h\rangle$. In order to describe how it transforms under rotations, one needs to introduce a $3\times3$ symmetric polarization matrix $\eps_{ij}(q)$. In order to project down to only 2 helicities, one can demand that the polarization matrix is transverse-traceless
\beq
\eps_{ii}=q_i\,\eps_{ij}=0.
\eeq
 Note that these constraints are manifestly rotationally invariant, so we have cut down to the correct number of degrees of freedom, while maintaining our assumed space-time symmetry. The situation is analogous for massless spin 1, which can be described in a manifestly rotationally invariant way by a polarization vector $\eps_i$ that is transverse $q_i\,\eps_i=0$. Hence, unlike the Lorentz invariant case, there is no a priori reason to introduce gauge invariance into these descriptions. 

{\em Long Range Interactions}:---
We would like to build a theory that contains long range interactions. However, if we attempt to do so using the above spin 1 or spin 2 particles we encounter a problem. To see this, consider the propagator for these particles (we assume parity here for simplicity, but our results do not rely upon this)
\bea
G_{ij}(q)&=& i{\delta_{ij}-{q_iq_j\over|{\bf q}|^2}\over E^2-K_1({\bf q})}, \\
G_{ijkl}(q) &=& -{i\over2}{\left(\delta_{ij}-{q_iq_j\over|{\bf q}|^2}\right)\left(\delta_{kl}-{q_kq_l\over|{\bf q}|^2}\right)-(j\leftrightarrow k,l)\over E^2-K_2({\bf q})},
\,\,\,\,\,\, \eea
where the index structure enforces these to be transverse (and traceless for spin 2) and $K_{1,2}$ are dispersion relations for the spin 1 and spin 2 particles, respectively.

Now consider $2\to2$ scattering between some matter particles via the exchange of a single spin 1 or spin 2 particle. Since the propagator carries indices, we are required to contract with some vector current $J^i$ or matrix $\tau^{ij}$ for spin 1 and spin 2, respectively. This gives the following contribution to the action for the matter degrees of freedom from tree-level exchange, which we show now for the spin 1 case, and extend to spin 2 in the Appendix:
\beq
\Delta S=\!\int\! {d^4q\over(2\pi)^4}\!\left[\tilde{J}^i(q) {\delta_{ij}-{q_iq_j\over|{\bf q}|^2}\over \omega^2-K_1({\bf q})} \tilde{J}^{j*}(q)+{\tilde\rho(q)\tilde\rho^*(q)\over L_1({\bf q})}   \right],
\label{Spin1DeltaS}\eeq
where we have also allowed for the exchange of a non-dynamical scalar $\phi$ that mediates some type of Coulomb interaction between charge density $\rho$. This second term is included for two reasons: (i) it is compatible with rotation invariance and does not introduce any additional degrees of freedom, and (ii) the first term evidently cannot by itself mediate a Coulomb-like interaction since it involves the currents $J^i$ associated with moving particles, and hence the first term does not give rise to a force between static charges, while the second term can.

For any current $J^i$ that does not trivially vanish in the soft limit, the first term here is invariably non-local due to the $\sim q_iq_j/|{\bf q}|^2$ structure, whose Fourier transform is long ranged. The Coulomb interaction being associated with a non-dynamical field is evidently also non-local. Both lead to  instantaneous action at a distance.

{\em Local Interactions}:---
In this work, we will impose the most basic version of locality:  we demand our theories do not have instantaneous action at a distance. 
In order for this to be possible, we require that the non-locality in the above action cancels out. 

To do so, there must be a constitutive relationship shared between $J^i$ and $\rho$; this relationship must be linear since both terms are of the same order. One can check that the most general form allowed is
\beq
q_i\,\tilde{J}^i=M_1({\bf q})\,\omega\,\tilde\rho,
\label{constitutive}\eeq
where $M_1$ is some function that mixes the two fields together. This equation reduces to the familiar charge conservation equation if $M_1=1$, but is different otherwise. 

If we take the classical particle limit for the charge density $\rho({\bf x},t)=\sum_n\,\en\,\delta^3({\bf x}-{\bf x}_n(t))$, and vary the action with respect to ${\bf x}_n(t)$, we obtain the force applied to a test charge $\en$ at position ${\bf x}_n(t)$. Let us assume that for all times $t<0$ the charge and current densities vanish, but are suddenly non-zero at time $t=0$. In order to avoid instantaneous action at a distance, every derivative of ${d^p\over dt^p}{\bf x}_n(t=0)$ ($p\geq2$) must vanish when the test charge is separated from the rest of the charges. In order for this to occur, every coefficient of $1/\omega^p$ in (\ref{Spin1DeltaS}) must be a local function if we expand $\Delta S$ in inverse powers of $\omega$.

By using the constitutive relationship to relate the $\sim\tilde J^i(q)(q_iq_j/|{\bf q}|^2)\tilde J^{j*}(q)$ and $\sim\tilde\rho(q)\tilde\rho^*(q)$ terms, and the fact that only polynomials of ${\bf q}^2$ have local Fourier transforms (derivatives of delta functions), we find the following conditions must be satisfied by the functions $K_1,\,L_1,\,M_1$:
\bea
K_1({\bf q})&=&m_1^2\,\delta_{P_0,0}+|{\bf q}|^2 P_a(|{\bf q}|^2),\label{K1}\\
L_1({\bf q})^{-1}&=&{P_1\over|{\bf q}|^2}+P_b(|{\bf q}|^2),\label{L1}\\
M_1({\bf q})^2&=&P_1+|{\bf q}|^2P_c(|{\bf q}|^2),\label{M1}
\eea
where $P_a,\,P_b,\,P_c$ are polynomials in $|{\bf q}|^2$, $P_1$ is a constant, and $m_1$ is a mass that must vanish unless $P_1=0$.  If we compute the Coulomb interaction between a pair of charges by Fourier transforming $L_1({\bf q})^{-1}$ we obtain
\beq
V({\bf x})=e_1e_2\left({P_1\over 4\pi |{\bf x}|}+P_b(-\Delta)\,\delta^3({\bf x})\right).
\eeq
Hence in order to have a long range interaction we require $P_1\neq 0$. We assume this going forward and can set $P_1=1$. We then find the following properties: (i) there is an exact $\sim 1/r$ Coulomb potential at large distances (plus negligible contact interactions), (ii) the photon is necessarily gapless, and (iii) we can integrate up the constitutive relationship (\ref{constitutive}) over all space and find {\em global} (though usually not local) charge conservation. 

{\em Soft Gauge Invariance}:--- 
The above $2\to2$ exchange can be obtained from a complete theory that allows for external photons. By introducing creation and annihilation operators to describe multi-particle states and Fourier transforming to position space $(\eps_i({\bf q})\,\hat{a}_{\bf q}+\eps_i^*({\bf q})\,\hat{a}_{-{\bf q}}^\dagger)/\sqrt{2E_q}\to A_i({\bf x})$, we can construct the following Lagrangian density for spin 1 fields
\bea
\cL&=&{1\over 2}|\dot{\bf A}|^2-{1\over2}\nabla\times{\bf A}\cdot\mathcal{K}_1(-\Delta)\nabla\times{\bf A} +{\bf A}\cdot{\bf J}-\phi\,\rho\nonumber\\
&+&{1\over 2}\phi\,\LP_1(-\Delta)\phi+\dot{\bf A}\cdot\mathcal{M}_1(-\Delta)\nabla\phi +\lm(\nabla\!\cdot\!{\bf A})^2,
\eea
where $\mathcal{K}_1(-\Delta)\equiv K_1(-\Delta)/(-\Delta)$ and $\mathcal{M}_1(-\Delta)\equiv L_1(-\Delta)M_1(-\Delta)/(-\Delta)$. One can check that this gives precisely the exchange action in (\ref{Spin1DeltaS}). Note that we have included a Lagrange multiplier $\lm$ to project out the longitudinal mode of ${\bf A}$. 

Now locality restricts the form of $K_1,\,L_1,\,M_1$ to that given above in Eqs.~(\ref{K1}-\ref{M1}). If we focus on only the terms with the lowest number of spatial derivatives, we see that the kinetic term for the spin 1 field organizes into $\cL_2={1\over2}|\dot{\bf A}+\nabla\phi|^2-{c^2\over2}|\nabla\times{\bf A}|^2$
and the theory becomes endowed with a gauge invariance
\beq
A_\mu\equiv A_\mu+\partial_\mu\alpha\,\,\,\,\,(\mbox{slowly varying $\alpha$})
\eeq
(combined with familiar gauge transformations for the matter sector)
with $A_\mu\equiv(-\phi,{\bf A})$ for {\em slowly varying} $\alpha$, which we refer to as `soft gauge invariance'. 

A similar result holds for spin 2 particles, where we can again Fourier transform $(\eps_{ij}({\bf q})\,\hat{a}_{\bf q}+\eps_{ij}^*({\bf q})\,\hat{a}_{-{\bf q}}^\dagger)/\sqrt{2E_q}\to h_{ij}({\bf x})$. In this case we need to firstly introduce a non-dynamical scalar $h_{00}$ to mediate a long ranged Newtonian force, and secondly we need a non-dynamical vector $h_{0i}$. Together they allow the theory to avoid instantaneous action at a distance when introduced appropriately, with similar properties to the above: (i) an exact $\sim 1/r$ Newtonian potential at large distances and (ii) a gapless graviton with $K_2({\bf q})=c_g^2|{\bf q}|^2+\ldots$. Moreover, all are organized into a $4\times4$ symmetric matrix $h_{\mu\nu}$, again endowed with a soft gauge invariance (see the Appendix)
\beq
h_{\mu\nu}\equiv h_{\mu\nu}+\partial_\mu\alpha_\nu+\partial_\nu\alpha_\mu\,\,\,\,\,(\mbox{slowly varying $\alpha_\mu$})
\label{soft2}\eeq
 
Intuitively, the presence of the soft gauge invariance is required to ensure that the non-dynamical fields are mixed with the propagating degrees of freedom such that long range forces inherit the finite speed of propagation of the spin 2 or spin 1 particles.

{\em Spin 1 Interacting Theories}:---
For spin 1, we can construct various Lorentz violating, but local Lagrangians. If we truncate to just two derivatives, then in fact we obtain exact gauge invariance. As an example, consider the following Lagrangian for massless spin 1 coupled to multiple species of fermions
\beq
\cL=-{1\over4}\eta^{\mu\alpha}\,\eta^{\nu\beta}\,F_{\mu\nu}F_{\alpha\beta} 
+ \sum_n \bar\psi_n (i\,\gamma_n^\mu D_\mu-m_n)\psi_n,
\label{QED}\eeq
where $D_\mu\psi_n=\partial_\mu\psi_n-i\,\en\,A_\mu\,\psi_n$ is the usual covariant derivative and the matrices $\{\gamma_n^\mu,\gamma_n^\nu\}=2\,\etan^{\mu\nu}$ encode arbitrary limiting speeds of propagation $c_n$ as follows:
\beq 
\etan^{\mu\nu}=\mbox{diag}(1,-c_n^2,-c_n^2,-c_n^2),
\label{QEDmatrix}\eeq
which violate Lorentz (but not gauge) invariance.

We note that this Lorentz violating construction carries over immediately to Yang-Mills fields by simply dressing up the spin 1 gauge fields with color indices and introducing a Lie algebra structure to encode self interactions in the usual way. Furthermore, this can be used to introduce various forms of Lorentz violation in the Standard Model at the dimension 4 level \cite{Colladay:1998fq,Coleman:1998ti}.

{\em Spin 2 Interacting Theories}:---
For spin 2, it is significantly more complicated to construct consistent theories. One possibility is to simply couple the linearized Riemann tensor $R_{\mu\nu\alpha\beta}$ directly to matter \cite{Wald:1986bj,Hertzberg:2016djj}. These models typically have problems with causality \cite{Hertzberg:2016djj,Hertzberg:2017abn}, but in any case do not mediate long range forces, and will not be further explored here. Instead we would like to explore leading order interactions. We know that one can introduce a consistent theory if one assumes the Lorentz symmetry, as this leads to general relativity. But, if we do not assume the Lorentz symmetry, then there are many challenges and open questions:
(i) Does the propagation speed of the matter sector need to agree with the propagation speed of the graviton? 
(ii) What restrictions are placed on a general dispersion relation $E_n^2=\tilde{K}_n({\bf p}_n)$ for the matter particles?
(iii) What changes occur if we include various types of matter species; fermions and gauge/vector bosons?
(iv) What constraints apply to the graviton's dispersion relation?
(v) What possible structures could we utilize to build a consistent theory, including conserved currents?
(vi) Is the equivalence principle still required for consistency?

In order to address these questions in a systematic fashion, we now utilize Weinberg's technique of studying a generic scattering process involving the emission of a soft graviton \cite{Weinberg:1964ew}.

\begin{figure}[t]
\hspace{1.1cm}\includegraphics[width=6cm]{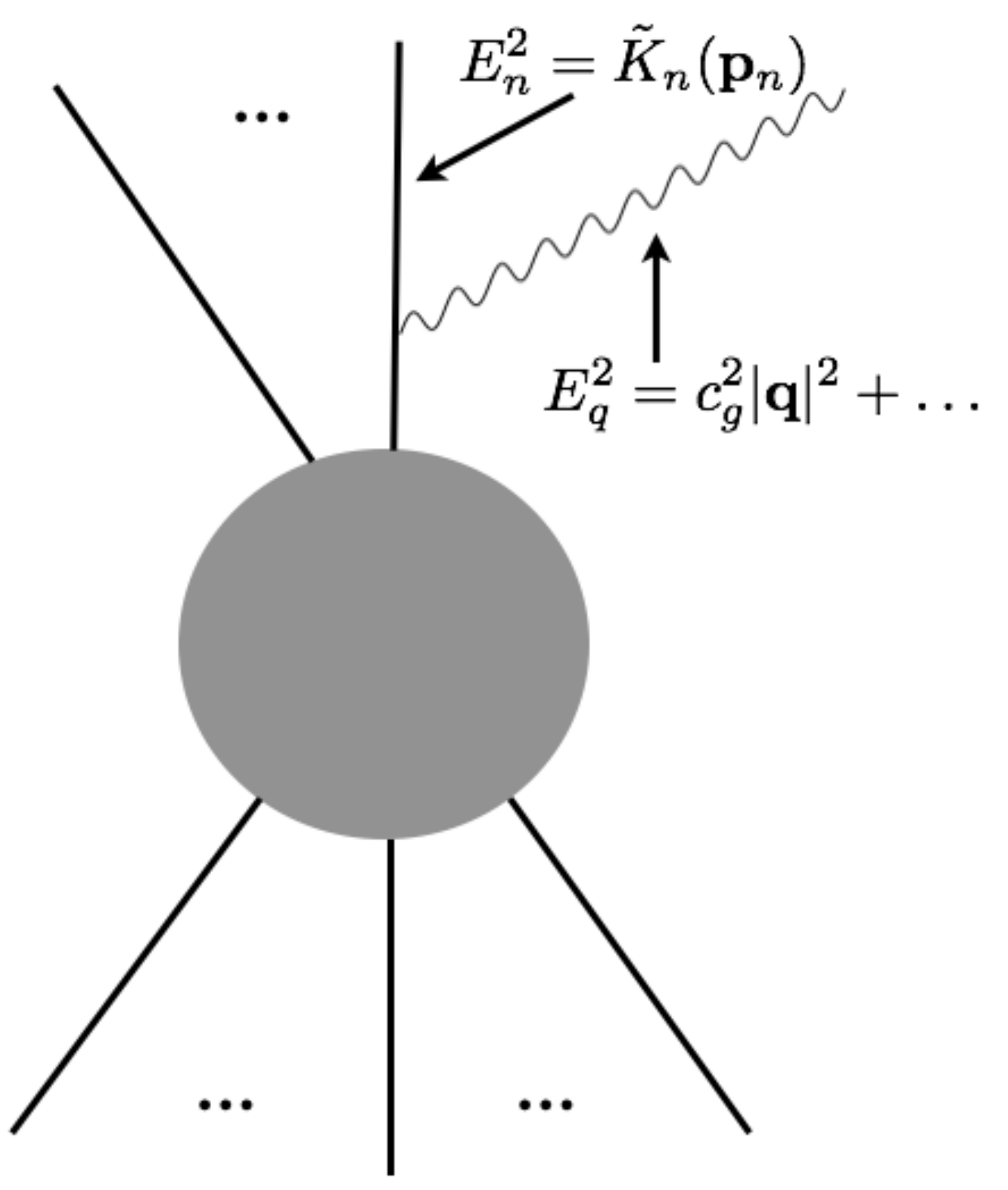}
\caption{General scattering process involving $N$ external particles (solid lines) with general dispersion relations $E_n^2=\tilde{K}_n({\bf p}_n)$ and a soft graviton (wiggly line) with infrared dispersion relation $E_q^2=c_g^2|{\bf q}|^2+\ldots$.}
\label{Figure1}\end{figure}

{\em Soft Graviton Emission}:---
Consider a general scattering processes involving $N+1$ external particles including a graviton. The amplitude $\cM_{N+1}$ is known to simplify in the soft graviton limit; namely it {\em factorizes} into the amplitude $\cM_N$ for $N$ particles times a piece describing the emission of the graviton from an external leg, as depicted in Figure \ref{Figure1}, as follows:
\bea
\cM_{N+1}=\cM_N\times&&\Bigg{[}\sum_{i}{\epsilon_{\mu\nu}(q)\,\ta_i^{\mu\nu}(p_i)\over(E_i-E_q)^2-\tilde{K}_i({\bf p}_i-{\bf q})}\nonumber\\
&&+\sum_f{\epsilon_{\mu\nu}(q)\,\ta_f^{\mu\nu}(p_f)\over(E_f+E_q)^2-\tilde{K}_f({\bf p}_f+{\bf q})}\Bigg{]},\,\,\,\,\,\,\,\,
\label{amplitude}\eea
where $\epsilon_{\mu\nu}(q)$ is the polarization matrix of the outgoing soft graviton, $\ta_n^{\mu\nu}(p_n)$ is some matrix associated with the $n$th matter particle that the graviton is coupled to, the $i$ subscript refers to initial particles, and the $f$ subscript refers to final particles. Note that we have taken each interaction vertex to involve the same virtual and external particles. If this were not the case and the two particles had different dispersion relations, then the virtual particle would be highly off-shell and would not contribute appreciably to the above sum in the soft limit.

{\em Constraint from Locality and Unitarity}:---
Under a soft gauge transformation the polarization matrix of the external graviton transforms as
\beq
\epsilon_{\mu\nu}(q)\to\epsilon_{\mu\nu}(q)+ q_\mu\,\tilde\alpha_\nu+ q_\nu\,\tilde\alpha_\mu\,\,\,\,\,(\mbox{soft} \,\,\tilde\alpha_\mu)
\eeq
where $q_\mu=(E_q,-{\bf q})$ is the graviton's 4-momenta. In order for the scattering amplitude to be associated with a local and unitary theory, it must be left unchanged. This ensures that the longitudinal modes are consistently removed and not being sourced by $\ta^{\mu\nu}_n$.

Furthermore, in the soft limit the respective denominators of (\ref{amplitude}) can be simplified by Taylor expanding to $\mathcal{O}(E_q,{\bf q})$ in the graviton's momentum. The need for soft gauge invariance leads to the constraint
\beq
\sum_{i}{q_\mu\,\ta_i^{\mu\nu}(p_i)\over q_\alpha\,\zeta^\alpha_i(p_i)}
=\sum_{f}{q_\mu\,\ta_f^{\mu\nu}(p_f)\over q_\alpha\,\zeta^\alpha_f(p_f)},
\label{cons2}\eeq
where $\zeta_n^\alpha(p_n)$ is a type of ``momentum" with an upstairs index; more precisely it is
\beq
\zeta_n^\alpha(p_n)\equiv\left(E_n,{1\over 2}{\partial\tilde{K}_n\over\partial{\bf p}}\Big{|}_{{\bf p}_n}\right)
\eeq

Now Eq.~(\ref{cons2}) must be valid for {\em any} graviton momentum, regardless of its direction and for any set of momenta for the particles. The only way this is possible is for the graviton momenta $q_\alpha$ to {\em cancel} out of numerator and denominator.  It is easy to see that this is only possible if the matrix $\ta^{\mu\nu}_n(p_n)$ that the graviton momenta is contracted with must be proportional to the $\zeta^\alpha_n$ vector in each of its indices. The most general form of $\ta^{\mu\nu}_n(p_n)$ is therefore
\beq
\ta_n^{\mu\nu}(p_n)=\g_n(E_n)\,\zeta^\mu_n(p_n)\,\zeta^\nu_n(p_n),
\eeq
where we have allowed for a prefactor $\g_n$ that can depend on any property of the $n$th particle that is a scalar under rotations, including energy. Inserting this into (\ref{cons2}) gives the conservation laws
\beq
\sum_i\,\g_i(E_i)\,\zeta^\nu_i(p_i)=\sum_f\,\g_f(E_f)\,\zeta^\nu_f(p_f),
\eeq

{\em Dispersion Relation}:---
The $\nu=0$ conservation law is $\sum_i\,\g_i(E_i)\,E_i=\sum_f\,\g_f(E_f)\,E_f$. Since the only conserved scalars are the energy itself and charges, the function $\g_n$ must take the form
\beq
\g_n(E_n)=\kappa+{\qn\over E_n},
\label{gsoln}\eeq
where $\kappa$ is a universal coupling and $\qn$ is some charge associated with the $n$th particle. The $\nu=i$ conservation law is
\beq
\sum_i \,g_i(E_i){\partial\tilde{K}_i\over\partial{\bf p}}\Big{|}_{{\bf p}_i}=
\sum_f \,g_f(E_f){\partial\tilde{K}_f\over\partial{\bf p}}\Big{|}_{{\bf p}_f}.
\eeq
Now the only conserved 3-vectors are the 3-momentum and angular momentum of the particles. But since $\partial\tilde K/\partial{\bf p}$ must be parallel to the 3-momentum, then this is the only viable option. Hence we have
\beq
g_n(E_n){\partial\tilde{K}_n\over\partial{\bf p}}\Big{|}_{{\bf p}_n}=\al\,{\bf p}_n ,
\eeq
where $\al$ is some universal coefficient. Using (\ref{gsoln}) and the fact that on-shell $\tilde{K}_n=E_n^2$, we can integrate up this equation to give
\beq
\kappa\,E_n^2+2\,\qn\,E_n = {\al\over2}|{\bf p}_n|^2+\ba_n,
\label{dispersion}\eeq
where $\ba_n$ is some constant of integration.

{\em Galilean Symmetry?}:---
To explore this dispersion relation, let us begin by considering the special case in which $\kappa=0$. By dividing throughout by the charge $\qn$ we obtain 
\beq
E_n={|{\bf p}_n|^2\over 2\,M_n}+\tilde\ba_n,
\label{Galilean}\eeq
where $M_n\equiv 2\,\qn/\al$ and $\tilde{\ba}_n\equiv\ba/(2\,\qn)$. Hence we discover the dispersion relation of Newtonian mechanics, and since the coupling $g_n\propto M_n$, we see a connection to Newtonian gravity. Furthermore we see that the mass of particles is required to be conserved as it is the re-scaled conserved charge $M_n\propto \qn$. So in fact (\ref{Galilean}) is the most general structure compatible with Galilean symmetry.

{\em Lorentz Symmetry}:---
The above $\kappa=0$ option that leads to Galilean symmetry among the matter sector is only a viable option if one restricts attention to processes in which {\em all} gravitons are taken to zero momentum. However, we are also allowed to consider hard gravitons as part of the $N$ particles participating in the scattering process, in addition to the one soft graviton. Since gravitons exchange momentum with the rest of the particles, they are required to be part of the momentum conservation law, and hence the dispersion relation (\ref{dispersion}) must apply to gravitons too. Since the gravitons have $E_q^2=\cg^2|{\bf q}|^2+\ldots$ at low momenta, then self consistency demands that 
\beq
\kappa\neq0\,\,\,\,\,\,\,\,\mbox{and}\,\,\,\,\,\,\,\,\al=2\,\kappa\,\cg^2
\eeq
and $Q_g=\ba_g=0$ for the graviton. We then discover that not only does the graviton's dispersion relation start linear for small momenta, it is required to stay {\em exactly} linear for all momenta.

Furthermore, for any particles that carry non-zero charge $\qn$, we can always absorb the charge into the definition of the energy by the replacement
\beq
E_n\to E_n-{\qn\over\kappa},
\eeq
since this maps a conserved energy into another conserved energy. Then without loss of generality the dispersion relation for {\em all} particles can be put into the form
\beq
E_n^2=|{\bf p}_n|^2\,c_g^2+m_n^2\,c_g^4.
\eeq
So we find the graviton's speed $c_g$ sets a universal speed limit, and we can replace $c_g\to c$. Hence we discover the complete and most general energy-momentum relation allowed is that of the familiar special relativistic form.

Then by noting that energy is the generator of time translations and momentum is the generator of spatial translations, we readily obtain that $ds^2\equiv c^2\,dt^2-|d{\bf x}|^2$ is invariant. This provides the hyperbolic structure of Minkowski space, boost invariance, and the Lorentz transformations. This goes a long way toward constructing special relativity from the ground up.

{\em Soft Photon Emission}:---
One may compare the above analysis to theories only involving particles with spin $s\leq1$. 
It is well known that with a massless spin 1 particle, which we shall refer to as a ``photon", requiring consistent soft scattering is still highly constraining
 \cite{Weinberg:1964ew}.

If we consider a soft outgoing massless spin 1 particle with matter obeying an arbitrary dispersion relation, then the matrix element is still  given by (\ref{amplitude}) with the replacement $\epsilon_{\mu\nu}(q)\,\ta_n^{\mu\nu}(p_n)\to \epsilon_\mu(q)\,\cJ^\mu_n(p_n)$. Here $\epsilon_\mu(q)$ is the polarization vector of the outgoing soft photon and $\cJ_n(p_n)$ is some vector associated with the $n$th matter particle. The locality and unitarity constraint for the soft photon 
\beq
\epsilon_\mu(q)\to\epsilon_\mu(q)+q_\mu\,\tilde\alpha \,\,\,\,\,(\mbox{soft} \,\,\tilde\alpha)
\eeq
then leads to a version of (\ref{cons2}) with $\ta_n^{\mu\nu}(p_n)\to \cJ^\mu_n(p_n)$ in the numerator, namely
\beq
\sum_{i}{q_\mu\,\cJ_i^{\mu}(p_i)\over q_\alpha\,\zeta^\alpha_i(p_i)}
=\sum_{f}{q_\mu\,\cJ_f^{\mu}(p_f)\over q_\alpha\,\zeta^\alpha_f(p_f)}.
\label{cons2photon}\eeq
In order for this to be satisfied for {\em any} photon and matter momenta, we need $q_\mu$ to cancel out, which requires
\beq
\cJ_n^\mu(p_n)=f_n(E_n)\,\zeta^\mu(p_n),
\eeq
where again we have allowed for a prefactor $f_n$ that is some scalar under rotations. Inserting this into (\ref{cons2photon}) gives the single conservation law
$\sum_i\,f_i(E_i)=\sum_f\,f_f(E_f)$. Again using the fact that the only conserved scalars are charges and energy, we have the general solution
\beq
f_n(E_n)=\en+{E_n\over M},
\eeq
where $\en$ is a charge and $M$ is some universal mass scale. 

In the language of Lagrangians, these terms are associated with the interaction terms
\beq
\Delta
\cL_{int}=A_\mu\,J^\mu+{1\over M}A_\mu\,T^{\mu}_{\,\,0},
\eeq
where the first term is the familiar coupling of the photon to a current and the second (Lorentz violating) term couples the photon to the time-components of the energy-momentum tensor. We do not know if the second term possesses a non-linear completion, but we shall not focus on this term here. What is most important is that the dispersion relation has completely dropped out of this analysis, and hence one can couple (at least using the $A_\mu\,J^\mu$ term) a massless photon to any Lorentz violating $\tilde{K}_n$ (an example was seen earlier in Eqs.~(\ref{QED}-\ref{QEDmatrix})). On the other hand, this is impossible when spin 2 is included.

{\em Discussion}:---
An interesting suggestion made in \cite{Khoury:2013oqa} is that the Lorentz symmetry might emerge as an accidental symmetry associated with massless spin 2 particles. In that work, only the graviton with only a 2 derivative action was studied. In this work, we have shown that even with an arbitrary number of matter species and higher derivatives, the relativistic dispersion relation must be exact in the IR and UV, rather than merely accidental.

It remains to be proven if all higher order interaction terms with matter must necessarily obey the Lorentz symmetry and to construct the entire Einstein-Hilbert action with matter species, but we believe it is impressive that the leading order interactions and the free sector are forced to carry the Lorentz symmetry exactly. 

Moreover, there are plausible reasons why typical interactions may be Lorentz invariant: (i) Suppose we consider a bosonic field that acquires some condensate configuration, plus small fluctuations $\phi=\phi_c+\delta\phi$. Then consider configurations where the length and time scale over which the condensate changes is slow compared to the soft graviton. Then from our work, the fluctuations should be found to be described by a Lorentz invariant theory, as they look like free fields at the quadratic order. This usually only arises from a fully Lorentz invariant interaction term to begin with. (ii) Non-Lorentz invariant interaction terms may renormalize lower dimension operators and be inconsistent with the relativistic dispersion relation derived here. (iii) It is possible that any Lorentz violating process is merely spontaneous breakdown.

Finally, our result that any local and unitary effective theory (in a universe like ours that includes gravity) must obey the special relativistic dispersion relation exactly, brings into serious question the viability of various proposed modifications of special relativity. Some examples include (i) the Lorentz violating construction of the Standard Model \cite{Colladay:1998fq,Coleman:1998ti}, (ii) Lorentz violating constructions of quantum gravity, such as Horava-Lifschitz \cite{Horava:2009uw}, (iii) the construction of various deformed versions of relativity \cite{AmelinoCamelia:2000mn,Magueijo:2001cr}, and (iv) some alternatives to inflation that appeal to an arbitrarily widened light cone at high energies \cite{Moffat:1992ud}.

{\em Acknowledgments}:---
We would like to thank Jose Blanco-Pilado, John Donoghue, Gary Goldstein, Ali Masoumi, Mohammad Namjoo, and Masaki Yamada for useful discussions, and the Tufts Institute of Cosmology.

{\em Email Correspondence}:--- 
mark.hertzberg@tufts.edu, mccullen.sandora@tufts.edu

\newpage

{\em Appendix}:--
Here we extend the locality analysis to the spin 2 case.  The most general interaction from tree-level graviton exchange is
\bea
\Delta S&=&\!\int\! {d^4q\over(2\pi)^4}\!\bigg[\tilde{\tau}_{ij} {\Proj_{ik}\Proj_{jl}-\frac12\Proj_{ij}\Proj_{kl}\over \omega^2-K_2({\bf q})} \tilde{\tau}_{kl}^* 
+\frac{2\,\tilde \pi_i\,\tilde\pi_i^*}{N_2(\bf q)}\nn\\
&&\,\,\,\,\,\,\,\,\,\,\,\,\,\,\,\,\,\,-{\tilde\md\,\tilde\tau_{ii}^*\over R_2({\bf q})}
-\frac12{\tilde\md\,\tilde\md^*\over L_2({\bf q})}-\frac12{\omega^2\tilde\md\,\tilde\md^*\over |{\bf q}|^2 L_2'({\bf q})}
  \bigg],\,\,\,\,\,\,
\label{Spin2DeltaS}\eea
where $\Proj_{ij}\equiv\delta_{ij}-{q_iq_j\over|{\bf q}|^2}$. For non-localities to cancel, we need the constitutive relations: $q_i\,\tilde{\tau}_{ij}=M_2({\bf q})\,\omega\,\tilde \pi_i$ and $q_i\,\tilde{\pi}_i=M_2'({\bf q})\,\omega\,\tilde\md$. Imposing locality we find that these functions must be related to polynomials as:
\bea
K_2({\bf q})&=&m_2^2\,\delta_{P_2P_2',0}+c_g^2|{\bf q}|^2+|{\bf q}|^4 P_d(|{\bf q}|^2),\label{K2}\\
N_2({\bf q})^{-1}&=&{P_2\over|{\bf q}|^2}+P_e(|{\bf q}|^2),\\
R_2({\bf q})^{-1}&=&{\sqrt{P_2P_2'}\over|{\bf q}|^2}+P_f(|{\bf q}|^2),\\
L_2({\bf q})^{-1}&=&{c_g^2P_2P_2'\over|{\bf q}|^2}+P_g(|{\bf q}|^2),\\
L_2'({\bf q})^{-1}&=&{P_2P_2'\over|{\bf q}|^2}+P_h(|{\bf q}|^2),\\
M_2({\bf q})^2&=&P_2+|{\bf q}|^2P_i(|{\bf q}|^2),\\
M_2'({\bf q})^2&=&P_2'+|{\bf q}|^2P_j(|{\bf q}|^2).
\eea
So a long range force requires $c_g^2P_2 P_2'\neq0$. Hence the graviton must be massless, and we can set $P_2=P_2'=1$, and we need $c_g\neq0$.  By Fourier transforming to the local field representation $(\eps_{ij}({\bf q})\,\hat{a}_{\bf q}+\eps_{ij}^*({\bf q})\,\hat{a}_{-{\bf q}}^\dagger)/\sqrt{2E_q}\to h_{ij}({\bf x})$, we can construct the following Lagrangian:
\bea
\cL&=&{1\over 2}|\dot{\bf h}|^2-{1\over2}\nabla\times{\bf h}\cdot\mathcal{K}_2(-\Delta)\nabla\times{\bf h} +h_{ij}\,\tau^{ij}+2\,\psi_i\,\pi_i\nonumber\\
&+&\phi\,\md+{1\over 2}\nabla\psi\cdot \mathcal{N}_2(-\Delta)\nabla\psi+\nabla\psi\cdot \mathcal{M}_2(-\Delta)\dot {\bf h} \nonumber\\
&+&\phi \,\mathcal{R}_2(-\Delta)\nabla\nabla\!\cdot\!\bf h+\lm(\nabla\!\cdot\!{\bf h})^2.
\label{Lgrav}\eea
For ease of notation, we have defined the dot product between two matrices as $A\!\cdot\! B\equiv A_{ij}B_{ij}-A_{ii}B_{jj}$.  The functions here are related to the above as $\mathcal{K}_2(-\Delta)\equiv K_2(-\Delta)/(-\Delta)$, $\mathcal{N}_2(-\Delta)\equiv N_2(-\Delta)/(-\Delta)$, $\mathcal{M}_2(-\Delta)\equiv M_2'(-\Delta)R_2(-\Delta)/(-\Delta)$, and $\mathcal{R}_2(-\Delta)\equiv R_2(-\Delta)/(-\Delta)$.  Demanding that the spin 2 exchange arises from an action places further consistency conditions on the functions $L_2^{-1}=K_2/R_2^2$, $L_2'^{-1}=4M_2'^2/N_2-3|{\bf q}|^2/R_2^2$, and $M_2=M_2'R_2/N_2$. Note that in our convention, the gravitational couplings are included in the sources $\tau^{ij},\,\pi^i,\,\md$.

We then find that to leading order in a derivative expansion, the first, second, sixth and seventh terms in (\ref{Lgrav}) assemble into $\cL_2={1\over2}|\dot{\bf h}+\nabla\psi|^2-{c_g^2\over2}|\nabla\times{\bf h}|^2$, which is invariant under the gauge transformation $h_{ij}\rightarrow h_{ij}+\nabla_{(i}\alpha_{j)}$, $\psi_i\rightarrow\psi_i-\dot\alpha_i$.  Likewise, the seventh and eighth terms in (\ref{Lgrav}) are invariant under the gauge transformation $\psi_i\rightarrow\psi_i-\nabla_i\alpha_0$, $\phi\rightarrow\phi+\dot\alpha_0$ (up to a total derivative).  A $4\times4$ matrix $h_{\mu\nu}$ can then be assembled as $h_{0i}=-\psi_i$, $h_{00}=\phi$, and we obtain soft gauge invariance, as reported in Eq. (\ref{soft2}).

\end{document}